\documentclass[11pt, a4paper,  notitlepage]{article}
 \usepackage{a4}
 \usepackage{epsfig}
 \usepackage{latexsym}

\author{B~Khoruzhenko$^\dag$
\\SFB 237 ``Unordung und gro\ss e Fluktuationen'' , Institut f\"ur Mathematik,
\\
Ruhr-Universit\"at Bochum,  44780 Bochum, Germany }
\title{
Large-$N$ Eigenvalue Distribution of Randomly Perturbed Asymmetric
Matrices
}

\begin{document}

\date{ }

\maketitle

\vskip 9cm

\noindent Short title: Eigenvalue Distibution of Random Asymmetric Matrices

\noindent PACS number(s): 05.40.+j

\vskip 2cm

\noindent $\phantom {l}^\dag$ On leave from: Institute for  Low
Temperature Physics, 310164, Kharkov, Ukraine

\newpage

\begin{abstract}
The density of complex eigenvalues of random asymmetric $N\times N$
matrices is found in the large-$N$ limit. The matrices are of the form
$H_0+A$ where $A$ is a matrix of $N^2$ independent, identically distributed
random variables with zero mean and variance $N^{-1}v^2$. The limiting
density $\rho (z,z^*)$ is bounded. The area of the support of
$\rho (z,z^*)$ cannot be less than  $\pi v^2$.
In the case of  $H_0$ commuting with its conjugate,
$\rho (z,z^*)$ is  expressed in terms of the eigenvalue distribution of
the non-perturbed part $H_0$.
\end{abstract}

\newpage

Random Hermitean and real symmetric matrices have been extensively
studied since the 50's, the time when Wigner introduced them into theoretical
physics.
A lot of results concerning these matrices and respective techniques
are known
now. In contrary to this, random complex and real asymmetric
matrices are much less studied. Although they have already
proved to be useful.
We mention here only two examples
(but see a discussion in \cite{Haake-1992}).
These are: i) quantum chaotic
scattering and decaying processes, where complex eigenvalues of
random non-Hermitean matrices are
used to analyse statistical properties of resonances \cite{ Haake-1992,
Sokolov-1988,  Lehmann-1995}, and ii) neural network dynamics where
synaptic matrices are in general asymmetric and the distribution of
their eigenvalues is  important for the understanding of network dynamics
\cite{Sompolinsky-1988, Doyon-1993}.

In this letter we consider random real
asymmetric matrices of the form
$H=H_0+A$. $A=[ a_{jk} ]_{j,k=1}^N$ is a matrix of $N^2$
independent, identically distributed random variables such that
\begin{eqnarray}
\langle a_{jk} \rangle =0,\;\; \langle a_{jk}a_{lm}\rangle =
N^{-1} v^2 \delta_{jl}\delta_{km}.
\label{A}
\end{eqnarray}
The angle brackets $\langle \ldots \rangle$ denote average over the
random variables $a_{jk}$.
For simplicity we assume that $a_{jk}$ are Gaussian but our
results remain valid for a wider class of distributions. We treat
$A$ as a perturbation and $H_0$ as the non-perturbed part and our aim is
to determine the large-$N$ limit of the averaged density
of complex eigenvalues of $H_0+A$.

If $A$ and $H_0$ are symmetric (or Hermitean) and $A$
obeys the GOE (GUE) statistics, then $H_0+A$ is known as
the deformed GOE (GUE)
\cite{Brody-1981}. In this ensemble eigenvalues are real and, hence,
their density is completely determined by the imaginary part of the
Green's function $G(E+i0)=
\langle N^{-1}\mathrm{tr}\, (E+i0-H)^{-1}\rangle$. It
appears that in the large-$N$ limit the Green's function of the
deformed GOE (GUE) is related to that of $H_0$ by the so-called Pastur's
equation \cite{Pastur-1972}:
\begin{eqnarray}
G(z)=G_0(z-v^2G(z)).
\label{Pastur}
\end{eqnarray}
Although this equation cannot be solved explicitly (except of a few cases)
it provides useful information about the density of eigenvalues.
For instance, one can prove that the density of
eigenvalues in the
deformed GOE (GUE) is bounded and generically decays as the square
root in the vicinity of the spectrum boundaries \cite{Khorunzhy-1992}.

Eigenvalues of asymmetric matrices are complex and their average
density $\rho (z,z^*)$  is determined by the electrostatic potential
\begin{eqnarray*}
\Phi (\kappa,z,z^*)=
- N^{-1}\langle \log \det [(zI-H)^*(zI-H)+\kappa^2I] \rangle
\end{eqnarray*}
by means of Poisson's equation $ \rho (z, z^*) =
-\frac{ \displaystyle 1 }{ \displaystyle \pi  }
\frac{ {\displaystyle \partial^2 \Phi (\kappa,z,z^*)} }
{ {\displaystyle \partial z \partial z^*} }\Bigl|_{\kappa =0}$
\cite{Haake-1992, Sommers-1988}. $I$ is the identity
matrix. Positive infinitesimal $\kappa$ is introduced in order to
regularize the potential. Provided $\kappa =0$, $\Phi$ as a function
of complex $z$ has a singularity whenever $z$ equals one of the
eigenvalues of $H$.

Anticipating an important role of positive
semi-definite matrices $\mathcal{H}=(zI-H)^*(zI-H)$ in studying
complex eigenvalues of $H$ we introduce the following Green's function
\begin{eqnarray}
R(\kappa )=\langle N^{-1}\mathrm{tr}\,(\mathcal{H}+\kappa^2 I)^{-1}
\rangle \,
\label{Green}
\end{eqnarray}
corresponding to $\mathcal{H}$. $R(\kappa )$ as a function of $\kappa $
is analytic in the right half of the complex plane and obviously
determines the density of eigenvalues of $\mathcal{H}$. We show that
in the large-$N$ limit the Green's functions of $\mathcal{H}$ and
$\mathcal{H}_0=(zI-H_0)^*(zI-H_0)$ are related by the equation (\ref{R})
which can be thought as generalization of Pastur's result to the case of
positive semi-definite random matrices. In passing we find derivatives
of the electrostatic potential. This allows us to derive an expression for
the average density of complex eigenvalues of $H$, $\rho (z, z^*)$, and
for the domain of their distribution. The respective expressions
(\ref{D})~-~(\ref{I}) are given in terms of $H_0$. Actually they set up
the only restriction to $H_0$: quantities entering  (\ref{D})~-~(\ref{I})
must be well defined in the large-$N$ limit. We do not specify $H_0$
further. It can be real or complex and either deterministic or random.
In the latter case it is assumed that $H_0$ is statistically independent
of $A$ and it is understood that the average over realizations of $H_0$
has been taken.

At this point it is worth mentioning that in the
specific case of $H_0$ commuting with its conjugate ${H_0}^*$ (i.~e. $H_0$
can be symmetric, skew-symmetric, Hermitean, skew-Hermitean, etc)
$\rho(z, z^*)$ can be {\sl explicitly} expressed in terms of the density
of eigenvalues of $H_0$ (see (\ref{dn})-(\ref{gn})). This should be compared
with the case of deformed GOE (GUE) where only the relation
(\ref{Pastur}) between Green's functions is known.

Our last remark
concerns matrices studied in \cite{Haake-1992,
Lehmann-1995}. They are of the form $iVV^\top +B$, where $V$ is an
$N\times M$ matrix of $NM$ independent Gaussian variables and
$B$ obeys the GOE statistics. These random matrices differ from
those considered here  in that aspect that $B$ is symmetric while $A$ is
asymmetric. The eigenvalue distribution of $VV^\top$ is known
\cite{Marchenko-1967} and it seems interesting to recover
the results of \cite{Haake-1992, Lehmann-1995},
which were obtained by means of the replica trick \cite{Haake-1992}
and supersymmetry calculations \cite{Lehmann-1995},
in the
framework of our approach. But this problem
goes beyond the aim of the present letter.

Introducing the notation $\mathcal{G}(\kappa )$ for the inverse
to $\mathcal{H} + \kappa^2 I$ we rewrite the following obvious matrix
identity $I=\langle\mathcal{G}(\kappa )(\mathcal{H} + \kappa^2
I)^{-1}\rangle$ as
\begin{eqnarray}
\kappa^2 \langle \mathcal{G}(\kappa) \rangle = I- (zI-H_0)^*
\langle(zI-H)\mathcal{G}(\kappa)\rangle + \langle A^*(zI-
H)\mathcal{G}(\kappa)\rangle\, .
\label{id}
\end{eqnarray}
$zI-H_0$ is statistically independent of $\mathcal{G}(\kappa)$
but $A$ which enters the $(zI-H)$ term in the r.h.s. of (\ref{id}) is not.
In order to decouple $\langle A\mathcal{G}(\kappa)\rangle$
and $\langle A^*A\mathcal{G}(\kappa)\rangle$ we first notice that
each of the entries $\mathcal{G}_{pq}$ of the matrix
$\mathcal{G}(\kappa)$ is a function of the Gaussian
variable $a_{lm}$. Therefore
\begin{eqnarray}
\langle a_{lm} \mathcal{G}_{pq} \rangle =
\langle a_{lm}^2 \rangle
\langle \partial \mathcal{G}_{pq}/\partial a_{lm} \rangle =
N^{-1}v^2 \,
\langle \partial \mathcal{G}_{pq}/\partial a_{lm} \rangle.
\label{Gauss}
\end{eqnarray}
This is the only place where Gaussian distribution of $a_{lm}$ is used.
In the non-Gaussian case it can be shown that
(\ref{Gauss}) holds up to the $1/N^2$ order if $a_{lm}=N^{-1/2}\alpha_{lm}$
and the random variables $\alpha_{lm}$
possess several first moments.
Straightforward application of (\ref{Gauss}) and the following rule for
differentiating
matrix elements of $\mathcal {G} (\kappa)$ with respect to those of~$A$
\begin{eqnarray}
\frac{\partial \mathcal{G}_{pq}}{\partial a_{lm}} =
[ \mathcal{G}(zI-H)^*]_{pl} \, \mathcal{G}_{mq} +
\mathcal{G}_{pm} \,
[ (zI-H)\mathcal{G} ]_{lq}.
\label{der}
\end{eqnarray}
gives
\begin{eqnarray*}
\langle (zI -H)\mathcal{G}(\kappa) \rangle =
(zI-H_0) \langle \mathcal{G}(\kappa) \rangle -
v^2\langle (zI-H) \mathcal{G}(\kappa)
\cdot N^{-1} \mathrm{tr} \, \mathcal{G}(\kappa) \rangle + O(1/N).
\end{eqnarray*}
One can readily check (\ref{der}) making use
of $\partial \mathcal{G}_{pq}
/\partial \mathcal{H}_{km} = -\mathcal{G}_{pk}\,\mathcal{G}_{mq}$
and of  the chain rule.
The normalized trace of $\mathcal{G}(\kappa )$ is a self-averaging
extensive quantity \cite{Khorunzhy-1992}. That is  it becomes
non-random in the large-$N$ limit:
$N^{-1}\mathrm{tr}\,\mathcal{G}(\kappa)= R(\kappa ) +O(1/N)$, where
$R(\kappa )=\langle N^{-1}\mathrm{tr}\,\mathcal{G}(\kappa)\rangle$.
Therefore we
conclude that
\begin{eqnarray}
\langle (zI-H)\mathcal{G}(\kappa) \rangle =(zI-H_0)\langle
\mathcal{G}(\kappa)\rangle
[1+v^2R(\kappa)]^{-1} + O(1/N).
\label{AG}
\end{eqnarray}
Similar reasoning leads to
\begin{eqnarray}
\langle A^*(zI-H)\mathcal{G}(\kappa) \rangle = -v^2 \kappa^2 R(\kappa)
\langle \mathcal{G}(\kappa ) \rangle + O(1/N).
\label{AAG}
\end{eqnarray}
Collecting (\ref{id}) and (\ref{AG})-(\ref{AAG}) we find that in
the leading order
\begin{eqnarray}
\langle \mathcal{G}(\kappa) \rangle =
\frac{1+v^2R(\kappa)}{(zI-H_0)^*(zI-H_0)+\kappa^2 [ 1+v^2R(\kappa)]^2I}\; .
\label{G}
\end{eqnarray}
Introducing the notation $\mathcal{G}_0(\kappa)$ for the inverse
to $\mathcal{H}_0 + \kappa^2I$ one can write (\ref{G}) in the form
$\langle \mathcal{G}(\kappa) \rangle=(1+v^2R(\kappa))G_0(\kappa
[1+v^2R(\kappa)] )$.
$R(\kappa )$ is to be determined from the self-consistency equation
\begin{eqnarray}
R(\kappa) = [1+v^2R(\kappa)] R_0(\kappa [ 1+v^2R(\kappa)] ),
\label{R}
\end{eqnarray}
where $R_0(\kappa)= N^{-1}\mathrm {tr}\,\mathcal{G}_0(\kappa )$.

Since
$-\partial \Phi /\partial z^* = \langle
N^{-1}\mathrm{tr} (zI-H)\mathcal{G}(\kappa) \rangle$ one can use
(\ref{AG}) and (\ref{G}) to calculate $\rho(z,z^*)$. Indeed,
\begin{eqnarray*}
-\frac{ \partial \Phi (\kappa,z,z^*) }
{ \partial z^* }=
N^{-1}\mathrm{tr}(zI-H_0)\mathcal{G}_0(\kappa [1+v^2R(\kappa )])
+O(1/N)
\end{eqnarray*}
Simple analysis of (\ref{R}) shows that in the leading order $ \frac{
\displaystyle \partial \Phi (\kappa,z,z^*) }
{\displaystyle \partial z^* } \Bigl|_{\kappa =0}$ in given by
\begin{eqnarray}
-\frac{ \partial \Phi (\kappa,z,z^*) }
{\partial z^* }
\Bigl|_{\kappa =0} =
N^{-1} \mathrm{tr}(zI-H_0)\mathcal{G}_0(\gamma (z,z^*)) ,
\label{deriv}
\end{eqnarray}
where $\gamma (z,z^*)=\lim_{\kappa \to 0+} \kappa [1+v^2R(\kappa )]$
is the solution of $R_0(\gamma ) = v^{-2}$
if $z$ lies inside the domain $D$ determined by the inequality
\begin{eqnarray}
R_0(0)= N^{-1}\mathrm{tr}[(zI-H_0)^*(zI-H)]^{-1}
\ge v^{-2}
\label{D}
\end{eqnarray}
and  $\gamma (z,z^*)=0$ otherwise.
Since in the latter case
$\frac{ {\displaystyle \partial \Phi} }
{ {\displaystyle \partial z^*} }\Bigl|_{\kappa =0}$ does not
depend on $z$ we conclude immediately that $\rho (z,z^*)=0$ outside $D$.
On the other hand, differentiating (\ref{deriv}) with respect to $z$
one finds that inside $D$
\begin{eqnarray}
\rho (z,z^*)=(\pi v^2)^{-1}- {\pi }^{-1}I(z,z^*)
\label{rho}
\end{eqnarray}
where $I(z,z^*)$ is the large-$N$ limit of
\begin{eqnarray}
\lefteqn{
N^{-1}\mathrm{tr}\,(zI-H_0)\mathcal{G}_0(\gamma (z,z^*))
(zI-H_0)^*\mathcal{G}_0(\gamma (z,z^*))}\nonumber \\
& - &
\left| N^{-1}\mathrm{tr}\,(zI-H_0)\mathcal{G}_0^2(\gamma (z,z^*))
\right|^2
\left[ N^{-1}\mathrm{tr}\,\mathcal{G}_0^2(\gamma (z,z^*))
\right]^{-1}.
\label{I}
\end{eqnarray}
For any two matrices $P$ and $Q$
$|\mathrm{tr} PQ^*|^2\le\mathrm{tr} \,PP^*\mathrm{tr} QQ^*$.
Therefore $I(z,z^*)\ge0$ and
$\rho (z,z^*)$ is bounded by $(\pi v^2)^{-1}$. This fact which is
interesting in its own has also an important consequence: the area of
$D$, the support of $\rho (z, z^*)$,
is not less  than  the area of a disk with radius $v$.

In order to illustrate the formulas derived we consider a few examples.
If $H_0=0$, then the equation $R_0(0)=v^{-2}$
which determines the boundary of $D$ takes the form $|z|^2=v^2$ and
$I(z,z^*)$ obviously vanishes. Thus, we recover the circular distribution
\cite{Ginibre-1965, Girko-1985,
Lehmann-1991}:
$\rho (z,z^*)$ equals $(\pi v^2)^{-1}$ inside
the disk $|z|\le v$ and zero outside.

\begin{figure}[hbtp]
\mbox{\epsfxsize=14cm\epsfbox{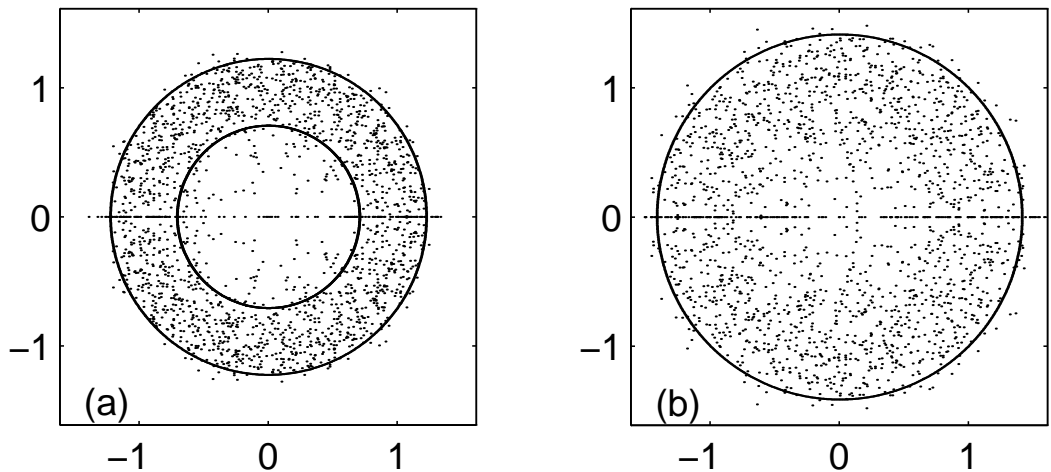}}
\caption{
Distribution of numerically computed eigenvalues of the random matrices  $J+A$
in the complex plane $z/h_0$.
In each of the plots $N=50$ and the number of samples is 40.  (a) $v^2/h_0^2
=1/2$,
 (b)  $v^2/h_0^2 =1$. The full lines show the boundary of  the support of $\rho
(z,z^*)$ in the
large-$N$ limit.
}
\label{fig1}
\end{figure}

In our next example $H_0$ is the Jordan block $J=[h_0 \delta_{j-
1,k}]_{j,k=1}^N$, $h_0>0$. $J$ has only one eigenvalue $z=0$ which is
defective
and highly sensitive to perturbations. On replacing zero in the
left lower corner of $J$ by small positive $\varepsilon$, one gets $N$
distinct eigenvalues $h_0(\varepsilon /h_0)^{1/N}\exp{(2\pi i k/N)}$.
For fixed $N$ the perturbed
eigenvalues approach zero as the parameter $\varepsilon /h_0$ vanishes
but the rate of
convergence is extremely slow if $N$ is large. For instance if $N=50$
one needs $\varepsilon /h_0 \propto 10^{-50}$ in order to confine the
eigenvalues into the disk $|z|/h_0 \le 0.1$. Therefore, if not exponentially
small, perturbation splits zero eigenvalue of $J$ into the circle
$|z|=h_0$. As can be seen from (\ref{D}) this phenomenon manifests
itself in the large-$N$ limit. Indeed, in  the case of $H_0=J$
(\ref{D}) reduces to $||z|^2-{h_0}^2|\le v^2$. Therefore if $v<h_0$  the
eigenvalues of $J+A$ are distributed in the annulus
$1-
\frac{ \displaystyle v^2  }
{ \displaystyle {h_0}^2  }
\le
\biggl| \frac{\displaystyle z }{\displaystyle h_0 }\biggr|^2 \le 1 +
\frac{\displaystyle v^2 }{\displaystyle {h_0}^2 }$
which degenerates into a circle
as $v$ vanishes.  When $v \ge h_0$ the eigenvalues
are distributed in the disk $
\biggl| \frac{\displaystyle z }{\displaystyle h_0 }\biggr|^2  \le 1 +
\frac{\displaystyle v^2 }{\displaystyle {h_0}^2 }$.
In Fig.\ref{fig1}  we present results of numerical diagonalization of random
matrices $J+A$. As can be seen, the correspondence between the
numerical results
and our analytical predictions (for $N \to \infty$) is quite good.

If $H_0$ commutes with its conjugate ${H_0}^*$, our formulas
(\ref{D})-(\ref{I})
become simpler. Let us assume for certainty that the eigenvalues of $H_0$
are real.  Then
the boundary of $D$ is determined by
\begin{eqnarray}
\int \frac{n(\lambda )d\lambda }{|z-\lambda|^2} = \frac{1}{v^2}\, ,
\label{dn}
\end{eqnarray}
where  $n(\lambda )$ is the  density of eigenvalues of $H_0$.
$\rho (z,z^*)$ is given by the same
expression (\ref{rho}) as before but now $I(z,z^*)$ is
\begin{eqnarray}
\lefteqn{
\int \frac{|z-\lambda|^2 n(\lambda )d\lambda}
{[|z-\lambda|^2 +\gamma^2(z,z^*)]^2}
}\nonumber \\
& - &
\left|\int \frac{(z-\lambda) n(\lambda )d\lambda}
{[|z-\lambda|^2 +\gamma^2(z,z^*)]^2}
\right|^2
\left[\int \frac{n(\lambda )d\lambda}
{[|z-\lambda|^2 +\gamma^2(z,z^*)]^2}
\right]^{-1}
\label{in}
\end{eqnarray}
and $\gamma (z,z^*)$ has to be found  from
\begin{eqnarray}
\int \frac{n(\lambda )d\lambda}{|z-\lambda|^2+\gamma^2} = \frac{1}{v^2}
\label{gn}
\end{eqnarray}

\vskip 1cm
This work was supported in part by the Deutsche Forschungsgemeinschaft
under Grant No SFB237 and by the International Science Foundation under
Grant No U2S000.

\end{document}